\documentclass[a4paper,twoside,titlepage,12pt]{article}

\usepackage{amssymb,amsmath,amsfonts}
\usepackage{epsfig}

\begin{document}

\begin{titlepage}

\rightline{NBI-HE-00-28}
\rightline{hep-th/0006023}
\rightline{June, 2000}
\vskip 2cm

\centerline{\Large \bf Supergravity and}
\vskip 0.2cm
\centerline{\Large \bf Space-Time Non-Commutative}
\vskip 0.2cm
\centerline{\Large \bf Open String Theory}
\vskip 0.2cm

\vskip 1.7cm
\centerline{{\bf Troels Harmark}\footnote{e-mail: harmark@nbi.dk} }
\vskip 0.3cm
\centerline{\sl The Niels Bohr Institute}
\centerline{\sl Blegdamsvej 17, DK-2100 Copenhagen \O, Denmark}
\vskip 2cm
\centerline{\bf Abstract}
\vskip 0.4cm

\noindent

We study the non-critical space-time non-commutative open string  
(NCOS) theory using a dual supergravity description in terms of 
a certain near-horizon limit of the F1-D$p$ bound state.
We find the thermodynamics of NCOS theory from supergravity. 
The thermodynamics is equivalent to Yang-Mills theory 
on a commutative space-time.
We argue that this fact does not have to be in contradiction with the
expected Hagedorn behaviour of NCOS theory. 
To support this we consider string corrections to the thermodynamics.
We also discuss the relation to Little String Theory in 6 dimensions.

\end{titlepage}


\newcommand{\nn}{\nonumber}
\newcommand{\spa}{\ \ ,\ \ \ \ }
\newcommand{\str}{\mathop{{\rm Str}}}
\newcommand{\tr}{\mathop{{\rm Tr}}}
\newcommand{\sn}{\mathop{{\rm sn}}}

\newcommand{\gym}{g_{\mathrm{YM}}}
\newcommand{\geff}{g_{\mathrm{eff}}}
\newcommand{\gseff}{g_s^{\mathrm{eff}}}
\newcommand{\Ord}{{\cal{O}}}
\newcommand{\tlst}{T_{\rm LST}}


\setcounter{page}{1}

\section{Introduction}

String theory in the presence of D$p$-branes 
with a magnetic NSNS $B$-field flux gives
a $p+1$ dimensional Yang-Mills theory 
on a non-commutative space (NCYM) in a 
certain decoupling limit \cite{Connes:1998cr,Seiberg:1999vs}.
Specifically, the coordinates of NCYM has the commutator
\begin{equation}
\label{com1}
[ x^i, x^j ] = i \theta^{ij} 
\end{equation}
Recently, it has been shown that one also can obtain a theory
with space and time non-commuting from string theory 
\cite{Seiberg:2000ms,Gopakumar:2000na}%
\footnote{See \cite{Seiberg:2000gc,Barbon:2000sg}
for other recent work on space-time
non-commutativity in string theory.}.
Thus, it has the commutator
\begin{equation}
\label{com2}
[ x^0 , x^i ] = i \theta^{0i}
\end{equation}
The new $p+1$ dimensional theory with this commutator is obtained
from the decoupling limit of D$p$-branes in the presence of a near-critical 
electrical NSNS $B$-field \cite{Seiberg:2000ms,Gopakumar:2000na}.
The theory thus obtained is not a field theory but a $p+1$ dimensional 
space-time non-commutative open string (NCOS) theory 
\cite{Seiberg:2000ms,Gopakumar:2000na}%
\footnote{That the theory with space and time 
non-commuting is not a field theory
could seem surprising since \eqref{com2} could be viewed
as the Lorentz-invariant completion of \eqref{com1}. 
But this is explained by the fact that 
field theories with space and time being non-commutative
are not unitary and thus cannot be a 
consistent truncation of string theory\cite{Gomis:2000zz}.}. 
NCOS theory is a new non-critical supersymmetric string 
theory which is believed to be non-gravitational and not to have 
a closed string sector \cite{Seiberg:2000ms,Gopakumar:2000na}%
\footnote{See \cite{Chen:2000ny} for other recent work on NCOS theory.}.

In this paper we study NCOS theory from supergravity.
Our starting point is the conjecture 
that $p+1$ dimensional NCOS theory is dual to 
string theory on the background given by 
a certain near-horizon limit of the F1-D$p$ 
bound state \cite{Gopakumar:2000na}%
\footnote{This is in the spirit of the near-horizon-D$p$-brane/QFT
correspondence \cite{Maldacena:1997re,Itzhaki:1998dd,Aharony:1999ti}}.
In \cite{Gopakumar:2000na} the near-horizon limit of the
F1-D3 bound state was constructed, and we generalize 
this to non-extremal F1-D$p$ bound states.
This adds to the list of correspondences
between string theory in the presence of D$p$-branes with a non-zero
NSNS $B$-field and dual non-gravitational theories, which has been
studied for magnetic NSNS $B$-field backgrounds in 
\cite{Maldacena:1999mh,Hashimoto:1999ut,Li:1999am,Alishahiha:1999ci,Harmark:1999rb,Lu:1999rm,Cai:2000hn}.

We use the near-horizon background to find the thermodynamics of NCOS theory
and we show that it is equivalent to the thermodynamics of 
ordinary Yang-Mills (OYM) theory, that is, Yang-Mills theory on a commutative
space-time.
The main goal of this paper is thus to understand how this thermodynamics
can be consistent with the expectation that NCOS theory is a string theory
and therefore should have Hagedorn behaviour.

We find that the thermodynamics obtained from supergravity
can be consistent with Hagedorn behaviour of NCOS theory. 
We find a region where the NCOS string coupling is very weak
and where the temperature is of the same order as the 
NCOS Hagedorn temperature, and by analyzing 
the phases of the supergravity description we find that this region 
cannot be described by supergravity. 
In other words, we find that one cannot both be near the Hagedorn 
temperature and have arbitrarily weak coupling 
in the supergravity description of NCOS theory.

By analyzing the high energy behaviour of the NCOS supergravity description
we find that it can be described by delocalized F-strings. 
This we use to approach the region mentioned above by doing tree-level
string corrections to the thermodynamics, and we find that
in this region NCOS theory has thermodynamics different from that of OYM.
This supports our conclusions that the supergravity thermodynamics
is consistent with NCOS Hagedorn behaviour.

We also consider 6 dimensional NCOS theory and the relation 
to space-time non-commutative Little String Theory. 
This is interesting since this gives us two
non-critical string theories which should be related to each other.

This paper is organized as follows. 
We give the 
non-extremal F1-D$p$ bound state in Section \ref{secF1Dp}.
We then take the near-horizon limit in Section \ref{secNHlimit}
and thereby obtain the background solution
that corresponds to finite temperature NCOS theory.
The phase structure of the supergravity description is studied in
Section \ref{secphases} and the NCOS thermodynamics is found and
discussed in Section \ref{secthermo}. 
We then show explicitly that the near-horizon background dual to NCOS theory
reduces to delocalized F-strings in Section \ref{secdelF1}.
This is used in Section \ref{secnewtherm} to search for a region
with new thermodynamics of the 4 dimensional NCOS theory.
We end by discussing the 6 dimensional NCOS theory and its relation to
Little String Theory in Section \ref{secp5}.

\section{The non-extremal F1-D$p$ bound state}
\label{secF1Dp}

In this section we present the non-extremal F1-D$p$ bound state
along with its thermodynamics%
\footnote{Non-extremal generalizations of non-threshold 
brane solutions were first considered in \cite{Costa:1996re}.}. 
In the extremal limit
it reduces to the extremal F1-D$p$ bound state 
given in \cite{Green:1996vh,Russo:1997if,Costa:1996zd,Lu:1999uc}.

The non-extremal F1-D3 bound state can be obtained by
S-duality from the non-extremal D1-D3 bound state%
\footnote{The non-extremal D1-D3 bound state can be found
in \cite{Harmark:1999rb} in a similar notation as this paper. 
One can then do the S-duality transformation 
\( g^E_{\mu \nu} \rightarrow g^E_{\mu \nu} \), 
\( e^{\phi} \rightarrow e^{-\phi} \), 
\( B_{\mu \nu} \rightarrow A_{\mu \nu} \),
\( A_{\mu \nu} \rightarrow - B_{\mu \nu} \), 
\( A_{\mu \nu \rho \sigma} \rightarrow A_{\mu \nu \rho \sigma} \) 
on the D1-D3 bound state where \( g^E_{\mu \nu} \) refers to the
Einstein-frame metric.}.
By use of T-duality on the non-extremal 
F1-D3 bound state we obtain
the non-extremal F1-D$p$ bound state with the string frame metric
\begin{eqnarray}
\label{NEmet}
ds^2 &=& \hat{D}^{-1/2} \hat{H}^{-1/2} \Big[ - f dt^2 + (dx^1)^2 \Big]
+ \hat{D}^{1/2} \hat{H}^{-1/2} \Big[ (dx^2)^2 + \cdots + (dx^p)^2 \Big]
\nn \\ && + \hat{D}^{-1/2} \hat{H}^{1/2} 
\Big[ f^{-1} dr^2 + r^2 d\Omega_{8-p}^2 \Big]
\end{eqnarray}
the dilaton
\begin{equation}
\label{NEdil}
e^{2\phi} = \hat{D}^{\frac{p-5}{2}} \hat{H}^{\frac{3-p}{2}}
\end{equation}
and the potentials
\begin{eqnarray}
\label{NEpot1}
B_{t1} &=& \sin \hat{\theta} ( \hat{H}^{-1} - 1 ) \coth \hat{\alpha}
\\
A_{2 \cdots p} &=& (-1)^p \tan \hat{\theta} ( \hat{H}^{-1} \hat{D} - 1 )
\\
\label{NEpot3}
A_{t1 \cdots p} &=& (-1)^p \cos \hat{\theta} \hat{D} ( \hat{H}^{-1} - 1 ) 
\coth \hat{\alpha}
\end{eqnarray}
where $B_{\mu \nu}$ is the NSNS two-form, and $A_{\mu_1 \cdots \mu_{p-1}}$
and $A_{\mu_1 \cdots \mu_{p+1}}$ are the RR $(p-1)$-form 
and $(p+1)$-form potentials\footnote{The RR five-form field strength
obtained from the RR four-form potential should be made self-dual.}.
We have also
\begin{equation}
\hat{H} = 1 + \frac{r_0^{7-p} \sinh^2 \hat{\alpha} }{r^{7-p}} \spa
f = 1 - \frac{r_0^{7-p}}{r^{7-p}}
\end{equation}
\begin{equation}
\hat{D}^{-1} = \cos^2 \hat{\theta} + \sin^2 \hat{\theta} \hat{H}^{-1}
\end{equation}
The thermodynamics is given by
\begin{equation}
\label{NEM}
M = \frac{V_p V(S^{8-p})}{16 \pi G} r_0^{7-p}
\Big[ 8-p + (7-p) \sinh^2 \hat{\alpha} \Big]
\end{equation}
\begin{equation}
\label{NETS}
T = \frac{7-p}{4 \pi r_0 \cosh \hat{\alpha} }
\spa
S =  \frac{V_p V(S^{8-p})}{4 G} r_0^{8-p} \cosh \hat{\alpha} 
\end{equation}
\begin{equation}
\label{NEmQF1}
\mu_{\rm F1} = -\sin \hat{\theta} \tanh \hat{\alpha}
\spa
Q_{\rm F1} 
= -\sin \hat{\theta} \frac{V_p V(S^{8-p})}{16 \pi G} (7-p) r_0^{7-p}
\cosh \hat{\alpha} \sinh \hat{\alpha}
\end{equation}
\begin{equation}
\label{NEmQDp}
\mu_{{\rm D}p} = \cos \hat{\theta} \tanh \hat{\alpha}
\spa
Q_{{\rm D}p} 
= \cos \hat{\theta} \frac{V_p V(S^{8-p})}{16 \pi G} (7-p) r_0^{7-p}
\cosh \hat{\alpha} \sinh \hat{\alpha}
\end{equation}
We note that just as for the D-brane bound state solutions 
the thermodynamics only depends on the angle $\hat{\theta}$ 
in the the charges and chemical potentials%
\footnote{See \cite{Harmark:1999rb} for an explicit account of this.}.
We also note that for the F1-D3 bound state we have the same
thermodynamics as the D1-D3 bound state which reflects the fact
that these two bound states are S-dual to each other.

Using charge quantization of the D$p$-brane we get
\begin{equation}
\label{chaqua}
r_0^{7-p} \cosh \hat{\alpha} \sinh \hat{\alpha} 
= \frac{(2\pi)^{7-p} g_s N l_s^{7-p}}{(7-p) V( S^{8-p} ) \cos \hat{\theta}}
\end{equation}
where $N$ is the number of coincident D$p$-branes.
The angle $\hat{\theta}$ is related to the number $M$ of F-strings
in the bound state as
\begin{equation}
\label{hattheta}
\tan \hat{\theta} = - \frac{Q_{\rm F1}}{Q_{{\rm D}p}} 
= \frac{V_1 T_{\rm F1} M}{V_p T_{{\rm D}p} N}
= g_s \frac{(2\pi l_s )^{p-1} }{V_{p-1}} \frac{M}{N}
\end{equation}
Here $V_p = \int dx^1 \cdots dx^p$, $V_1 = \int dx^1$ and 
$V_{p-1} = \int dx^2 \cdots dx^p$.
We now go to another choice of variables for the solution
\eqref{NEmet}-\eqref{NEpot3} which for our purposes are more natural.
We introduce the variables $\alpha$ and $\theta$ as
\begin{equation}
\label{newvar}
\sinh^2 \alpha = \cos^2 \hat{\theta} \sinh^2 \hat{\alpha} \spa
\cosh^2 \theta = \frac{1}{\cos^2 \hat{\theta}}
\end{equation}
In terms of the variables \eqref{newvar} 
we can write the metric, dilaton and NSNS $B$-field as
\begin{eqnarray}
\label{newmet}
ds^2 &=& H^{-1/2} \Big[ D \Big( - f dt^2 + (dx^1)^2 \Big)
+  (dx^2)^2 + \cdots + (dx^p)^2 \Big]
\nn \\ && + H^{1/2} \Big[ f^{-1} dr^2 + r^2 d\Omega_{8-p}^2 \Big]
\end{eqnarray}
\begin{equation}
\label{newdil}
e^{2\phi} = H^{\frac{3-p}{2}} D
\end{equation}
\begin{equation}
\label{newpot}
B_{t1} = \tanh \theta \sqrt{ 1 + \cosh^{-2} \theta \sinh^{-2} \alpha }
( D H^{-1} - 1 )
\end{equation}
with
\begin{equation}
\label{newHD}
H = 1 + \frac{r_0^{7-p} \sinh^2 \alpha}{r^{7-p}} \spa
D^{-1} = \cosh^2 \theta - \sinh^2 \theta H^{-1}
\end{equation}
We see that the solution \eqref{newmet}-\eqref{newHD} are similar in form
to that of the D$(p-2)$-D$p$ bound state with a NSNS $B$ field flux 
turned on, only here the $B$ field is turned on in an electrical
component rather than a magnetic component. 
For the extremal F1-D$p$ bound state one can in fact obtain
the solution by Wick-rotating the Lorentzian D$p$-brane solution
into an Euclidean solution and then use T-duality and rotations
in the same manner as for magnetic $B$ fields, in order to turn
an electrical $B$-field component on\cite{Maldacena:1999mh}. 
After Wick rotating back
to a Lorentzian solution one then has the F1-D$p$ bound state.
This, however, does not work in the more general case of a non-extremal
F1-D$p$ bound state. Here one cannot obtain the F1-D$p$ bound state
from the D$p$-brane by T-duality and rotations.
This is due to subtleties occuring when doing T-duality in an Euclidean 
time-direction\cite{Hull:1998vg}.

In the new variables \eqref{newvar} the thermodynamics
\eqref{NEM}-\eqref{NEmQDp} is given by
\begin{eqnarray}
\label{newM}
M &=& \frac{V_p V(S^{8-p})}{16 \pi G} r_0^{7-p}
\Big[ 8-p + (7-p) \cosh^2 \theta \sinh^2 \alpha \Big]
\\
T &=& \frac{7-p}{4 \pi r_0 \sqrt{ 1 + \cosh^2 \theta \sinh^2\alpha } }
\\
S &=& \frac{V_p V(S^{8-p})}{4 G} r_0^{8-p} \sqrt{ 1 + \cosh^2 \theta \sinh^2\alpha }
\\
\mu_{\rm F1} &=& - \frac{\sinh \theta \sinh \alpha}{\sqrt{1+\cosh^2 \theta \sinh^2 \alpha}}
\\
Q_{\rm F1} &=& - \frac{V_p V(S^{8-p})}{16 \pi G} (7-p) r_0^{7-p} \sinh \theta \sinh \alpha \sqrt{ 1+\cosh^2 \theta \sinh^2 \alpha}
\\
\mu_{{\rm D}p} &=& \frac{\sinh \alpha}{\sqrt{1+\cosh^2 \theta \sinh^2 \alpha}}
\\
\label{newQDp}
Q_{{\rm D}p} &=&  \frac{V_p V(S^{8-p})}{16 \pi G} (7-p) r_0^{7-p} \sinh \alpha \sqrt{ 1+\cosh^2 \theta \sinh^2 \alpha}
\end{eqnarray}
The charge quantization relation \eqref{chaqua} becomes
\begin{equation}
\label{newchaqua}
r_0^{7-p} \sinh \alpha \sqrt{ \sinh^2 \alpha + \cosh^{-2} \theta }  
= \frac{(2\pi)^{7-p} g_s N l_s^{7-p}}{(7-p) V( S^{8-p} ) \cosh \theta}
\end{equation}
%

\section{The NCOS near-horizon limit}
\label{secNHlimit}

In this section we apply the limit found in \cite{Gopakumar:2000na} to the 
non-extremal F1-D$p$ bound state.
This gives a dual string theory description of the $p+1$ dimensional NCOS
theory in terms of string theory on a curved background space-time in the
presence of a near-critical electrical NSNS $B$ field.
As explained in \cite{Seiberg:2000ms,Gopakumar:2000na} 
the near-critical electrical NSNS $B$ field 
gives dynamical open string modes in the decoupled theory
since the open string tension almost cancels with the $B$ field
giving a finite effective open string tension.

Following \cite{Gopakumar:2000na}, 
we take the limit \( l_s \rightarrow 0 \) keeping fixed
\begin{equation}
\label{lim1}
u = \frac{r}{l_s} \spa
u_0 = \frac{r_0}{l_s} \spa
b = l_s^2 \cosh \theta \spa
\alpha = \mbox{fixed} \spa
\tilde{g} = \frac{g_s l_s^2}{b}
\end{equation}
Moreover, we rescale the world-volume coordinates
\begin{equation}
\label{lim2}
\tilde{x}^i = \frac{l_s}{b} x^i\ , \ i=0,1 \spa
\tilde{x}^j = \frac{1}{l_s} x^j\ , \ i=2,...,p
\end{equation}
with the notation \( t = x^0 \).
Notice that we keep \( u = r / l_s \) finite since the mass of an
open string stretched between to D-branes with distance $r$
measured in units of \( l_s^{-1} \) is \( l_s M \sim u \). 
Thus, contrary to the usual scaling of the radial
coordinate \( u = r /l_s^2 \) we do not let the string modes measured in
units of \( l_s^{-1} \) grow to infinity, but instead keep them finite
and thereby dynamical.

We note that the rescaling of coordinates in \eqref{lim1} and \eqref{lim2} 
makes all coordinates dimensionless. This means that when measuring
distance or time on the world-volume with the coordinates
\( \tilde{t},\tilde{x}^1,...,\tilde{x}^p \) 
we are measuring in units of \( \sqrt{b} \), and similarly we
measure energy in units of \( 1/ \sqrt{b} \).

Defining
\begin{equation}
R^{7-p} \equiv  \frac{(2\pi)^{7-p}}{(7-p) V( S^{8-p} )} \tilde{g} N
\end{equation}
we have from \eqref{newchaqua} that
\begin{equation}
u_0^{7-p} \sinh^2 \alpha = R^{7-p}
\end{equation}
Using the near-horizon limit defined by \eqref{lim1}-\eqref{lim2}
on \eqref{newmet} and \eqref{newdil} we get the string-frame metric and dilaton
\begin{eqnarray}
\label{NCOSmet}
\frac{ds^2}{l_s^2} &=& H^{1/2} \frac{u^{7-p}}{R^{7-p}} 
\Big[ - f d\tilde{t}^2 + (d\tilde{x}^1)^2 \Big] 
+ H^{-1/2} \Big[ (d\tilde{x}^2)^2 + \cdots + (d\tilde{x}^p)^2 \Big]
\nn \\ &&
+ H^{1/2} \Big[ f^{-1} du^2 + u^2 d\Omega_{8-p}^2 \Big]
\end{eqnarray}
\begin{equation}
\label{NCOSdil}
g_s^2 e^{2\phi} 
= \tilde{g}^2 \left( 1 + \frac{u^{7-p}}{R^{7-p}} \right) H^{\frac{3-p}{2}}
\end{equation}
where
\begin{equation}
H = 1 + \frac{R^{7-p}}{u^{7-p}}
\end{equation}
From \eqref{newpot} we get the NSNS $B$-field 
\begin{equation}
\label{NCOSpot}
B_{t1} = l_s^2 \frac{u^{7-p}}{R^{7-p}}
\end{equation}
We note that the constant appearing in the NSNS $B$-field in \eqref{newpot}
has been gauged away before taking the limit.
This gives a non-zero electrical component
for the world-volume field strength $F_{01}$ on the D$p$-branes.
The limit $l_s \rightarrow 0$ keeping fixed 
\eqref{lim1} and \eqref{lim2} corresponds to
approaching the critical value of the electrical field $F_{01}$.

\section{Phases of the supergravity description}
\label{secphases}

We now analyze the phases of the supergravity near-horizon solution
\eqref{NCOSmet}-\eqref{NCOSpot}.

The F1-D$p$ bound state in the limit \eqref{lim1} and \eqref{lim2} 
is believed
to describe the $p+1$ dimensional NCOS theory. 
The open strings of NCOS theory live in a $p+1$ dimensional space-time with the
time $\tilde{t}$ and the space-direction $\tilde{x}^1$ being 
non-commutative\cite{Seiberg:2000ms,Gopakumar:2000na} 
with commutator $[\tilde{t},\tilde{x}^1]=i$. 

The string length in NCOS theory 
is $\sqrt{b}$ \cite{Seiberg:2000ms,Gopakumar:2000na}, so since 
we measure length
in units of $\sqrt{b}$ the NCOS string length in our supergravity 
description is 1.
Thus, if there is any Hagedorn behaviour of the thermodynamics
it should occur at $T \sim 1$.
The open string coupling constant of NCOS theory is \( G_{\rm o} \) which is related
to \( \tilde{g} \) as \( \tilde{g} = G_{\rm o}^2 \) 
\cite{Gopakumar:2000na}.
The rescaled radial parameter $u$ is the energy of the
brane probe in units of \( 1/\sqrt{b} \). Thus $u$ is the 
effective energy scale in NCOS theory.

It is easily seen that for \( u \ll R \) the near-horizon solution reduces
to that of the ordinary near-horizon limit of D$p$-branes
(see for example \cite{Itzhaki:1998dd}). Thus, for these 
energies the NCOS theory reduces to an ordinary Yang-Mills (OYM) theory
in a $p+1$ dimensional commutative space-time.
One can therefore regard OYM as a low energy effective theory for NCOS theory.
$u \sim R$ is the energy scale where
the non-commutativity start appearing.
Since NCOS theory reduces to OYM for \( u \ll R \) 
we define the effective coupling constant
\begin{equation}
\geff^2 = (2\pi)^{p-2} \tilde{g} N u^{p-3} 
\end{equation}
This coupling constant corresponds to the usual effective coupling 
constant for OYM defined in \cite{Itzhaki:1998dd}
with $\gym^2 = (2\pi)^{p-2} \tilde{g} $ the Yang-Mills coupling
constant and $N$ the rank of $U(N)$. 
Note that \( \geff^2 \sim R^{7-p} u^{p-3} \).

For \( u \gg R \) the near-horizon solution instead reduces to 
being the near-horizon limit of delocalized F-strings. This
will be further explained and studied in Section \ref{secdelF1}.
Since \( u \sim R \) is the point where we go from the 
near-horizon D$p$-brane phase to the near-horizon delocalized F-string phase,
we expect that this is the point where the non-commutative effects
become significant, in analogy with the near-horizon D$(p-2)$-D$p$ 
description of NCYM.

The curvature of \eqref{NCOSmet} in units of \( l_s \) is
\begin{equation}
\label{curv}
\mathcal{C} = \frac{1}{\sqrt{u^4 + R^{7-p} u^{p-3} }} 
\end{equation}
In order for the near-horizon 
solution \eqref{NCOSmet}-\eqref{NCOSpot}
to accurately describe NCOS theory we need both \( \mathcal{C} \ll 1 \)
and \( g_s e^{\phi} \ll 1 \).
We now analyze these conditions in the two cases \( u \ll R \)
and \( u \gg R \).

If we consider \( u \ll R \) where the dual theory is OYM, we see that
\( \mathcal{C} \ll 1 \) is equivalent to having \( \geff^2 \gg 1 \).
Thus, we have the usual demand that the effective coupling should
be large. 
The condition \( g_s e^\phi \ll 1 \) is equivalent to
\( \geff^2 \ll N^{\frac{4}{7-p}} \), thus we get
\begin{equation}
\label{stdcond}
1 \ll \geff^2 \ll N^{\frac{4}{7-p}}
\end{equation}
so that \( N \gg 1 \)
which means we are in the planar limit of OYM. The condition
\eqref{stdcond} corresponds to 
the standard supergravity description of OYM 
given in \cite{Maldacena:1997re,Itzhaki:1998dd}.
Combining the condition \( u \ll R \) with \eqref{stdcond}
we get the additional conditions
that \( \tilde{g} \gg 1/N \) for \( p > 3 \), 
\( 1/N \ll \tilde{g} \ll 1 \) for \( p=3 \) and
\( \tilde{g} \ll 1 \) for \( p < 3 \).

Considering instead \( u \gg R \) we see that \( g_s e^{\phi} \ll 1 \) if
\begin{equation}
\label{dilsmall}
u^{7-p} \ll \frac{R^{7-p}}{\tilde{g}^2} \sim \frac{N}{\tilde{g}}
\end{equation}
Combining this with \( u \gg R \) we get that \( \tilde{g} \ll 1 \) 
which means that the near-horizon solution describes
weakly coupled NCOS theory. 
The bound \( u^{7-p} \ll N/\tilde{g} \) means that for larger energies
the theory flows to the S-dual near-horizon brane description.
For $p=3$ we go to NCYM described by the near-horizon
D1-D3 solution \cite{Ganor:2000my,Gopakumar:2000na}.
We discuss the case $p=5$ in Section \ref{secp5}. 

Demanding \( \mathcal{C} \ll 1 \) for \( u \gg R \)
we see from \eqref{curv} that we need \( u \gg 1 \).
This we can split in two cases.
If \( R \gg 1 \), which is equivalent to \( \tilde{g} \gg 1/N \),
we only need to have \( u \gg R \).
If \( R \ll 1 \), which is equivalent to \( \tilde{g} \ll 1/N \),
we see that we need to have \( u \gg 1 \) instead.

Thus, if we want to consider NCOS theory with $p \geq 3$ when \( \tilde{g} \ll 1/N \)
we need to have \( u \gg 1 \). 
This means that there is an energy range \( R \ll u \ll 1 \)
in which we cannot describe NCOS theory with arbitrarily weak coupling.
This will be important for our discussion of thermodynamics in 
Section \ref{secthermo}.
For completeness we note that for $p<3$ we can also have
$\tilde{g} \ll 1/N$ if $u \ll (\tilde{g} N)^{1/(3-p)}$
which means that we are in the OYM region.

We can also ask if it possible to find a regime with finite $N$. 
From the previous paragraph we immediately get that this is
possible if and only if \( u \gg 1 \).

\section{NCOS thermodynamics}
\label{secthermo}

Using the limit \eqref{lim1}-\eqref{lim2}
we have from \eqref{newM}-\eqref{newQDp} the thermodynamics 
\begin{equation}
\label{NCOSTS}
T = \frac{7-p}{4\pi u_0} \left( \frac{u_0}{R} \right)^{\frac{7-p}{2}}
\spa
S = \frac{\tilde{V}_p V(S^{8-p}) }{32 \pi^6 \tilde{g}^2} u_0^{8-p} 
\left( \frac{R}{u_0} \right)^{\frac{7-p}{2}}
\end{equation}
\begin{equation}
\label{NCOSEF}
E = \frac{\tilde{V}_p V(S^{8-p}) }{128 \pi^7 \tilde{g}^2} 
\frac{9-p}{2} u_0^{7-p}
\spa
F = - \frac{\tilde{V}_p V(S^{8-p}) }{128 \pi^7 \tilde{g}^2} 
\frac{5-p}{2} u_0^{7-p}
\end{equation}
with 
\begin{equation}
\tilde{V}_p = b^{-1} l_s^{2-p} V_p
\end{equation}
being the world-volume in the rescaled coordinates given by \eqref{lim2}.
The energy $E$ is the energy above extremality 
$E = M - \sqrt{ Q_{\rm F1}^2 + Q_{{\rm D}p}^2 } $.
We note that in order to obtain \eqref{NCOSTS} and \eqref{NCOSEF} 
it is necessary to rescale $T$, $E$ and $F$ as
\begin{equation}
T \rightarrow \frac{b}{l_s} T
\spa
E \rightarrow \frac{b}{l_s} E
\spa 
F \rightarrow \frac{b}{l_s} F
\end{equation}
since we have rescaled the time as \( \tilde{t} = t l_s /b \).

The curvature in units of \( l_s \) at the horizon is
\begin{equation}
\label{epsD}
\varepsilon_D = \frac{1}{\sqrt{u_0^4 + R^{7-p} u_0^{3-p} }}
\end{equation}
The dilaton squared at the horizon is
\begin{equation}
\label{epsL}
\varepsilon_L = \tilde{g}^2 \left( 1 + \frac{u_0^{7-p}}{R^{7-p}} \right)
\left( 1 + \frac{R^{7-p}}{u_0^{7-p}} \right)^{\frac{3-p}{2}}
\end{equation}
Thus the thermodynamics \eqref{NCOSTS} and \eqref{NCOSEF} is
valid for NCOS theory when \( \varepsilon_D \ll 1 \) and \( \varepsilon_L \ll 1 \).
This gives the phase structure of Section \ref{secphases} 
with the only difference
that $u$ is replaced with $u_0$. 
We have defined $\varepsilon_D$ and $\varepsilon_L$ for later convenience.

The thermodynamics \eqref{NCOSTS} and \eqref{NCOSEF} is clearly equivalent%
\footnote{See e.g. Appendix A in \cite{Harmark:1999xt}.}
to that of OYM at strong 't Hooft coupling and
with the identification $\gym^2 = (2\pi)^{p-2} \tilde{g}$.
This is not surprising for energies \( u_0 \ll R \), since here
the near-horizon solution \eqref{NCOSmet}-\eqref{NCOSpot} 
is the usual near-horizon D$p$-brane solution.
For $u_0 \gg R$ we get instead a prediction of the thermodynamics of
NCOS theory. 

If we consider the specific case of $p=3$ we know that
for \( u_0^{7-p} \sim N/\tilde{g} \) we go to the NCYM description
which also has the thermodynamics \eqref{NCOSTS} and \eqref{NCOSEF}
\cite{Maldacena:1999mh}, so the thermodynamics is given by this whenever
a supergravity description is available.

The fact that the thermodynamics does not change when raising
the energy or temperature naturally raises the question 
about the stringy nature of NCOS theory. 
NCOS theory is claimed to be a string theory with a string spectrum and this
should give rise to Hagedorn behaviour of the thermodynamics.
Hagedorn behaviour occurs in a string theory for large temperatures
and energies. When doing statistical mechanics
of the string states the partition function can only be defined for
temperatures lower than a certain maximal temperature, called the 
Hagedorn temperature. Near this temperature the thermodynamics behave
in a singular manner.
Clearly, the fact that \eqref{NCOSTS} and \eqref{NCOSEF} behaves
uniformly for all energies means that there is no sign of any Hagedorn
behaviour in this thermodynamics. Assuming of course $p \neq 5$
since for $p=5$ the thermodynamics \eqref{NCOSTS} and \eqref{NCOSEF}
does exhibit Hagedorn behaviour, as we shall discuss further in 
Section \ref{secp5}.

The Hagedorn temperature of NCOS theory is of order \( T \sim 1 \)
since the string scale is equal to 1. 
Thus, we should first test whether the temperature 
given in \eqref{NCOSTS} can reach the Hagedorn temperature.
That $T \sim 1 $ is equivalent to 
\( u_0^{5-p} \sim R^{7-p} \), or, equivalently,
\begin{equation}
u_0^{5-p} \sim \tilde{g} N
\end{equation}
Thus, assuming $p < 5$, we can see that if \( \tilde{g} N \gg 1\)
we need \( u_0 \gg 1 \), and if \( \tilde{g} N \ll 1 \)
we need \( u_0 \ll 1 \).
But, we know from Section \ref{secphases} 
that if we want \( \tilde{g} \ll 1/N \)
we must have \( u_0 \gg 1 \), so that means that we cannot both
be near the Hagedorn temperature and have arbitrarily weak coupling.
This is a very important observation, since this can explain why
the thermodynamics \eqref{NCOSTS} and \eqref{NCOSEF} does not
exhibit Hagedorn behaviour. 
The explanation being that when \( \tilde{g} \gg 1/N \) strong coupling
effects removes the Hagedorn behaviour of the string theory. This
is perfectly possible in that the statistical derivation of  
Hagedorn behaviour assumes arbitrarily weak coupling.
On the other hand, when \( u_0 \gg 1 \) we can have arbitrarily
weak coupling \( \tilde{g} \ll 1/N \) but this means that
\( T \gg 1 \) so that we are far above the Hagedorn temperature.
Thus, the Hagedorn temperature is clearly not limiting and we have
a Hagedorn phase transition at \( T \sim 1 \).

The fact that we have the thermodynamics \eqref{NCOSTS} and
\eqref{NCOSEF} for \( \tilde{g} \gg 1/N \) and \( u_0 \gg R \)
suggests that the NCOS theory reduces to OYM in this region 
because of strong coupling effects.

In Section \ref{secnewtherm} we show that when approaching \( u_0 \sim 1 \)
for arbitrarily weak coupling new thermodynamics can occur. 
This supports our explanation of the missing Hagedorn behaviour
in the thermodynamics.

\section{NCOS theory and delocalized F-strings}
\label{secdelF1}

In this section we elaborate on the fact that for \( u \gg R \)
the NCOS near-horizon solution from Section \ref{secNHlimit} reduce to that
of the near-horizon limit of delocalized F-strings.

The solution for $M$ coincident F-strings delocalized in $p-1$ 
directions is
\begin{equation}
ds^2 = \hat{H}^{-1} \Big[ -f dt^2 + (dx^1)^2 \Big]
+ (dx^2)^2 + \cdots + (dx^p)^2 + f^{-1} dr^2 + r^2 d\Omega_{8-p}^2
\end{equation}
\begin{equation}
e^{2\phi} = \hat{H}^{-1}
\end{equation}
\begin{equation}
B_{t1} = (\hat{H}^{-1} -1 ) \coth \hat{\alpha}
\end{equation}
where we have used the functions and variables defined 
in Section \ref{secF1Dp}. This is the supergravity solution that the F1-D$p$
bound state reduces to when $\hat{\theta} = \pi/2$.
From charge quantization of the F-strings we get
\begin{equation}
r_0^{7-p} \cosh \hat{\alpha} \sinh \hat{\alpha}
= \frac{(2\pi)^6 g_s^2 l_s^6}{(7-p)V(S^{8-p})} \frac{M}{V_{p-1}} 
\end{equation}
This can also be gotten by combining \eqref{chaqua} and \eqref{hattheta}.

Taking the near-horizon limit \( l_s \rightarrow 0 \) with the
variables in \eqref{lim1} and \eqref{lim2} fixed we get the
near-horizon solution
\begin{equation}
\label{F1metNH}
\frac{ds^2}{l_s^2} = \frac{u^{7-p}}{R^{7-p}} 
\Big[ - f d\tilde{t}^2 + (d\tilde{x}^1)^2 \Big] 
+ (d\tilde{x}^2)^2 + \cdots + (d\tilde{x}^p)^2
+ f^{-1} du^2 + u^2 d\Omega_{8-p}^2
\end{equation}
\begin{equation}
\label{F1dilNH}
g_s^2 e^{2\phi} = \tilde{g}^2 \frac{u^{7-p}}{R^{7-p}}
\end{equation}
\begin{equation}
\label{F1potNH}
B_{t1} = l_s^2 \frac{u^{7-p}}{R^{7-p}}
\end{equation}
Thus we clearly see that the solution \eqref{NCOSmet}-\eqref{NCOSpot} 
reduces to the solution \eqref{F1metNH}-\eqref{F1potNH}
for \( u \gg R \).

The curvature of \eqref{F1metNH} and the dilaton \eqref{F1dilNH} are
small if and only we have
\begin{equation}
1 \ll u^{7-p} \ll \frac{N}{\tilde{g}}
\end{equation}
If \( \tilde{g} N \gg 1 \) we need in addition that 
\( u^{7-p} \gg \tilde{g} N \) in order to be describing NCOS theory.
Thus, we need that \( \tilde{g} \ll 1 \). 

From \eqref{hattheta} we get that 
\begin{equation}
\tilde{g} = \frac{\tilde{V}_{p-1}}{(2\pi)^{p-1}} \frac{N}{M}
\end{equation}
This is the connection between $M$ and $N$ which makes it possible
to describe NCOS theory in terms of $M$ delocalized F-strings.

In \cite{Lu:1999rm,Cai:2000hn,Youm:2000ub} 
D$(p-2)$-branes delocalized in two directions
where shown to describe $p+1$ dimensional 
NCYM with a rank 2 non-commutative space at high energies.
From this it was conjectured that the world-volume theory 
of the D$(p-2)$-brane, which is OYM with $p-2$ space dimensions,
is equivalent to the NCYM theory.
It would be interesting to investigate if it is possible
to make a similar connection for delocalized F-strings. 
The world-volume theory of $N$ coincident F-strings
is the free orbifold CFT which is S-dual to OYM in 1+1 dimensions
\cite{Dijkgraaf:1997vv}.

\section{Approaching new thermodynamics}
\label{secnewtherm}

In this section we go beyond the leading order supergravity solution
in order to find evidence that the thermodynamics of NCOS theory can differ
from that of OYM. 
We choose to consider the special case $p=3$ only, 
but a similar analysis can be made for other dimensions.

The leading order type IIB supergravity receives string corrections of
two types. Derivative corrections with expansion parameter 
\( \alpha' =l_s^2 \)
and loop corrections with expansion parameter \( g_s \). 
These two types of expansions can be translated into two types 
expansions of the thermodynamics of the 
near-horizon NCOS supergravity solution. 
Thus, the derivative expansion has the expansion parameter 
\( \varepsilon_D \) given in \eqref{epsD}, and the loop expansion
has expansion parameter \( \varepsilon_L \) given in \eqref{epsL}
\footnote{A general account of how this works for various 
near-horizon backgrounds can be found in \cite{Correia:2000}.}.
But, in order to use scaling arguments we need that the expansion
parameters can be expressed as a constant times a power of $u_0$.
Thus, we need either \( u_0 \ll R \) or \( u_0 \gg R \).

In type IIB string theory the first two corrections to the leading
order supergravity action 
is the $l_s^6 R^4$ term and the $g_s^2 l_s^6 R^4$
term. These will translate into a correction of order $\varepsilon_D^3$
and a correction of order $\varepsilon_L \varepsilon_D^3$.

If we consider the case \( u_0 \ll R \) which corresponds to
4 dimensional OYM we have 
\begin{equation}
\varepsilon_D \sim \frac{1}{\sqrt{\tilde{g}N}} \spa
\varepsilon_L = \tilde{g}^2
\end{equation}
Thus, we see that neither of the expansion parameters depends on
\( u_0 \). This means for the OYM that the thermodynamics essentially
stays the same when approaching \( \varepsilon_D \sim 1 \) or
\( \varepsilon_L \sim 1 \). Thus, we will keep having 
the free energy \( F \propto T^4 \) but the coefficient in front
will be renormalized%
\footnote{For OYM there is a factor $3/4$
in difference between the free theory and the strongly coupled theory
described by supergravity, as discussed in \cite{Gubser:1998nz}.}.

If we instead consider the case \( u_0 \gg R \) which corresponds
to 4 dimensional NCOS theory, we have
\begin{equation}
\varepsilon_D = \frac{1}{u_0^2} \spa
\varepsilon_L \sim \frac{\tilde{g}}{N} u_0^4 
\end{equation}
If we consider the transition point \( \varepsilon_L \sim 1 \),
or, equivalently, \( u_0^4 \sim N/\tilde{g} \) we should not
get any new thermodynamics since the theory should flow into
4 dimensional NCYM as described in \cite{Ganor:2000my,Gopakumar:2000na}. 
Since NCYM also has
\( F \propto T^4 \) the thermodynamics should be preserved when
approaching this transition point. And, indeed, it can be shown 
\cite{Correia:2000} that the \( \varepsilon_L \sim 1 \) transition
is very smooth due to the fact that the leading term that
contributes around this point is of order 
\( \varepsilon_D^3 \varepsilon_L \) which is highly suppressed since
\( \varepsilon_D \ll 1 \).

On the other hand, if we consider the transition point 
\( u_0 \sim 1 \), all the tree-level terms in type IIB string theory
contribute. They form together a series of terms in the expansion
parameter \( \varepsilon_D = 1/u_0^2 \), with the first term being
\( \varepsilon_D^3 = 1/u_0^6 \).
Therefore we get new thermodynamics when approaching \( u_0 \sim 1 \).
This requires in fact \( \tilde{g} \ll 1/N \) since we still want
\( u_0 \gg R \). 
This supports our arguments of Section \ref{secthermo} since we here
stated that the Hagedorn behaviour should be observed in the region
\( u_0 < 1 \) with very weak coupling \( \tilde{g} \ll 1/N \).
We now see that exactly when we approach this region we get
new thermodynamics.

From the arguments above, we note that 
the corrected free energy of 4 dimensional NCOS theory
for $T$ approaching \( 1/(\sqrt{\tilde{g}N}) \) becomes
\begin{equation}
F = - \frac{\pi^2}{8} N^2 \tilde{V}_3 T^4 \left[
1+ \sum_{n=3}^\infty \frac{a_n}{(\tilde{g} N)^n T^{2n}}  \right]
\end{equation}
where \( a_n \) are undetermined coefficients.

\section{The F1-D5 and D1-NS5 bound states}
\label{secp5}

We conclude this paper with a closer look at the $p=5$ case. 
This should correspond to NCOS theory in 6 dimensions, or, for low
energies, 6 dimensional OYM. 
But, it is well-known that the D5-brane and NS5-brane world-volume
theory is the so-called Little String Theory (LST) 
\cite{Seiberg:1997zk,Berkooz:1997cq,Losev:1997hx}, which
is a non-critical supersymmetric closed string theory in 6 dimensions.
This means that NCOS theory in 6 dimensions is related to space-time
non-commutative LST, and from this the question arises 
of how an open string theory can be related to a closed string theory
on a non-commutative space-time.
This question could be important for future research since we here
have two non-critical string theories which are claimed
to be descriptions of the same theory.

Our proposal for how NCOS theory and LST are related is that NCOS theory is a low energy
limit of space-time non-commutative LST, since when we reach 
\( u_0^2 \sim N/\tilde{g} \) the theory flow to the near-horizon
limit of the D1-NS5 bound state. 
Thus, we propose to define space-time non-commutative LST as the
decoupling limit of D1-NS5, just as spatially non-commutative LST
is defined as the decoupling limit of D2-NS5 or D3-NS5 
\cite{Alishahiha:2000er}.
And, we moreover propose that for sufficiently low energies this
theory reduces to that of 6 dimensional NCOS theory.

From \eqref{NCOSTS} we get the Hagedorn temperature
of the space-time non-commuta{-}tive LST
\begin{equation}
\tlst = \frac{1}{2\pi\sqrt{\tilde{g}N} }
\end{equation}
Thus, the LST Hagedorn temperature is different from that of NCOS theory,
which confirms our suspicion that we have two different string scales
in the theory.

Moreover, if we have \( \tilde{g} \ll 1/N \), which we expect is 
necessary in order to see the NCOS Hagedorn behaviour, we get
that \( \tlst \gg 1 \) so that any Hagedorn transition in NCOS theory
would happen at much lower temperatures than \( \tlst \).
If, on the other hand, \( \tilde{g} N \gg 1 \) we have that
\( \tlst \ll 1 \) which means that there cannot be any NCOS
Hagedorn transition since the LST Hagedorn temperature 
is limiting \cite{Harmark:2000hw,Berkooz:2000mz} and \( T \sim 1 \)
thus cannot be reached. 

The above consideration supports our proposal 
that NCOS theory is a low energy limit of space-time non-commutative LST since 
it means that whenever the NCOS Hagedorn temperature \( T_{\rm NCOS} \)
is defined we have that \( T_{\rm NCOS }  \ll \tlst \).

Note also that the inverse string tension in LST is 
\( \alpha'_{\rm LST} = \tilde{g} \), while in NCOS theory it is
\( \alpha'_{\rm NCOS} = 1 \).
Thus, the low energy limit of LST is the limit \( \tilde{g} \rightarrow 0 \)
which precisely is the limit where NCOS theory should be applicable since
it is weakly coupled. Also, for \( \tilde{g} \rightarrow 0 \)
we are in the F1-D5 description which corresponds to NCOS theory.

We now examine the LST Hagedorn behaviour of this theory using the arguments
given in \cite{Harmark:2000hw}. 
If we are at energies so that \( u_0 \gg R \) we have that 
\begin{equation}
\varepsilon_D^3 \varepsilon_L \sim \frac{\tilde{g}}{N} \frac{1}{u_0^2}
\end{equation}
which gives 
\begin{equation}
S(T) \propto \frac{1}{\tlst-T}
\end{equation}
This is valid for \( u_0^2 \ll N/\tilde{g} \).
For \( u_0^2 \gg N/\tilde{g} \) we flow into the near-horizon
limit of the D1-NS5 bound state.
If we raise the energy sufficiently, we end up with 
a solution of D-strings delocalized in 4 directions.
One can show that for these we get the thermodynamics
\begin{equation}
S(T) \propto \frac{1}{(\tlst - T )^{2/3}} 
\end{equation}
valid for all sufficiently high energies.
This is the same critical behaviour as for spatially non-commutative
LST \cite{Harmark:2000hw,Correia:2000}.
It would be interesting if one could reproduce this from 
statistical arguments 
like it was done in \cite{Harmark:2000hw,Berkooz:2000mz}
for LST on a commutative space-time.

\section{Discussion}

In this paper we have studied NCOS theory using the 
dual supergravity description.
We have presented the non-extremal F1-D$p$ bound state 
and its NCOS near-horizon limit.
We have found that the thermodynamics to leading order is 
equivalent to that of OYM.
We have furthermore argued that this does not have to be in contradiction
with the expected Hagedorn behaviour of NCOS theory.

By considering string corrections to delocalized F-strings we found
that the thermodynamics becomes very different from that of OYM
when approaching the region with $u_0 < 1$ and very weak coupling
$\tilde{g} \ll 1/N $. This supports our conclusion that the supergravity
thermodynamics is not in contradiction with the expectations from NCOS theory.
The string correction analysis is also important since it is quantitative 
evidence that the thermodynamics only is equivalent to
OYM at leading order
in the supergravity. When string corrections are included
the S-dual backgrounds of near-horizon D1-D3 and F1-D3 have 
different thermodynamics.

\section*{Acknowledgments}

We thank N. Obers for useful discussions.

\addcontentsline{toc}{section}{References}


\providecommand{\href}[2]{#2}\begingroup\raggedright\endgroup

\end{document}